# OSS-CRS: Liberating AIxCC Cyber Reasoning Systems for Real-World Open-Source Security

Andrew Chin[†] Dongkwan Kim[†] Yu-Fu Fu[†] Fabian Fleischer[†] Youngjoon Kim[†]
HyungSeok Han[‡] Cen Zhang[†] Brian Junekyu Lee[†] Hanqing Zhao[†] Taesoo Kim[†‡]
[†] *Georgia Institute of Technology*, [‡] *Microsoft*

*Abstract*—DARPA's AI Cyber Challenge (AIxCC) showed that cyber reasoning systems (CRSs) can go beyond vulnerability discovery to autonomously confirm and patch bugs: seven teams built such systems and open-sourced them after the competition. Yet all seven open-sourced CRSs remain largely unusable outside their original teams, each bound to the competition cloud infrastructure that no longer exists. We present OSS-CRS, an open, locally deployable framework for running and combining CRS techniques against real-world open-source projects, with budget-aware resource management. We ported the first-place system (ATLANTIS) and discovered 10 previously unknown bugs (three of high severity) across 8 OSS-Fuzz projects. OSS-CRS is publicly available.

## 1. Introduction

AI-generated vulnerability reports [4, 18, 36, 47] are increasing the burden on open-source maintainers. The curl project shut down its bug bounty program after AI-written submissions overwhelmed reviewers with unconfirmed findings [45, 46]; FFmpeg maintainers criticized Google for reporting valid bugs without providing patches, calling the reports "CVE slop" [42]. In both cases, AI automation stops at discovery, leaving maintainers to validate findings and write fixes.

DARPA's AI Cyber Challenge (AIxCC, 2023–2025) [15] demonstrated that the gap between vulnerability discovery and validated remediation can be closed. Seven finalist teams built autonomous cyber reasoning systems (CRSs) that go beyond discovery: each CRS dynamically confirms a vulnerability with a proof of vulnerability (PoV) and synthesizes a patch validated against that PoV [14]. After the competition, all seven teams open-sourced their systems, in principle making it possible to deploy this capability against any open-source project.

Yet over half a year later, these CRSs remain largely unusable outside their original teams: each system is bound to team-specific infrastructure and interfaces. This adoption gap blocks the communities that should benefit from these systems: researchers cannot run comparable cross-CRS experiments under consistent settings; CRS developers lack a stable integration contract; and security practitioners cannot adopt deployable, budget-aware workflows that produce actionable findings.

To understand what prevents adoption, we analyzed the open-sourced codebases of all seven AIxCC finalists and identify three deployment barriers: ❶ *Infrastructure duplication*, each team independently rebuilt the same platform services; ❷ *Cloud lock-in*, every system targets the competition's Azure and Kubernetes environment, which was decommissioned after the finals; and ❸ *Monolithic design*, analysis techniques are embedded in monolithic systems, preventing researchers from comparing or combining them across teams. Even ATLANTIS, the first-place system [24], requires over 20 Azure virtual machines and cannot target new projects without its original cloud environment.

To address these barriers, we present OSS-CRS [37], an open framework for developing, running, and composing CRSs on real-world open-source projects. OSS-CRS provides a shared infrastructure layer including LLM budget management and cross-CRS artifact exchange, so that CRS developers can focus on analysis logic rather than platform engineering. It adopts the OSS-Fuzz [3] project format as its target interface, enabling any integrated CRS to target over 1,000 OSS-Fuzz projects without per-project customization. To validate the framework, we ported ATLANTIS and ran it against 8 OSS-Fuzz projects, discovering 10 previously unknown bugs, including three of high severity. OSS-CRS and all integrated CRSs are publicly available.

This paper makes the following contributions:

- An *empirical analysis* of all seven AIxCC finalist CRS codebases, identifying three deployment barriers that prevent practical reuse: infrastructure duplication, cloud lock-in, and monolithic design (§2).
- OSS-CRS, an open-source framework that addresses these barriers through a unified execution model and standard interface, budget-aware resource management across CPU, memory, and LLM usage, and support for combining CRS techniques across systems (§3).
- *Real-world validation*: porting ATLANTIS to OSS-CRS and discovering 10 previously unknown bugs (three of high severity) across 8 OSS-Fuzz projects, showing that competition-grade CRS techniques can be deployed without cloud infrastructure (§4, §5).

## 2. Background and Motivation

### 2.1. Cyber Reasoning Systems

A *cyber reasoning system* (CRS) is an autonomous agent that discovers and repairs software vulnerabilities without human intervention. The concept originated in DARPA's Cyber Grand Challenge (CGC, 2014–2016) [13, 48], in which teams built systems to attack and defend custom binaries. AIxCC [15] extended the concept to real-world open-source software in C and Java, using challenge targets drawn from OSS-Fuzz projects [3].

**CRS capabilities.** Traditional security tools typically focus on a single stage of vulnerability management: fuzzers discover crashes, static analyzers flag potential flaws, and program repair systems synthesize patches, but integrated pipelines from discovery through a validated fix remain uncommon.

A CRS unifies these stages. *Bug finding* generates a proof of vulnerability (PoV), an input that triggers abnormal execution such as a crash or sanitizer violation. *Bug fixing* synthesizes a patch and validates it by rebuilding the target and rerunning tests, confirming that the PoV no longer triggers while preserving the program's original functionality. A CRS can accept a variety of inputs: an entire source tree, a commit-level code diff, a SARIF report [35] from a static analyzer, or a fuzzing seed corpus.

**Practical scenarios.** This end-to-end capability supports several deployment settings. In CI/CD, a pipeline submits a pull-request diff and receives a PoV or confirmation of no bugs through a stable, machine-readable interface. For security practitioners, the system turns static-analysis findings into validated patches while enforcing spending limits. For research, multiple CRSs run on the same source tree with matched inputs, so results are reproducible and directly comparable.

**Infrastructure requirements.** These scenarios create concrete system requirements. A CRS must coordinate multiple tools across stages, pass artifacts such as crash inputs, PoVs, candidate patches, and build outputs between them, and recover when a stage fails. LLM calls span bug finding, triage, and patch generation, thus token use must be budgeted and capped like CPU time and memory. AIxCC allocated $50,000 in LLM credits per team; without budget controls, a single CRS run can exceed $1,000 per hour [5]. Orchestration, artifact exchange, and budget-aware execution are therefore core design problems, not implementation details.

**OSS-Fuzz.** The AIxCC competition drew its targets from Google's OSS-Fuzz [3], which provides reproducible, containerized builds for over 1,000 open-source projects. Each project defines a Dockerfile and build script that compile the target project with sanitizer instrumentation, and supplies fuzz targets that consume fuzzer-generated inputs. For CRS deployment, this offers a standardized way to build and test many projects without per-project setup. However, OSS-Fuzz runs one fuzzer per container and does not provide multi-component orchestration, cross-stage artifact exchange, or budget management.

**TABLE 1:** Deployment characteristics of the AIxCC finalist CRSs. "Local": can run end-to-end on a single machine against a new target without cloud provisioning. "Composable": analysis components can be extracted and recombined across systems.

| CRS | Comp. | Infra | Middleware | Local | Composable |
|---|---|---|---|---|---|
| ATLANTIS [24] | 9+ | TF+K8s | Ka/PG+R* | × | × |
| BUTTERCUP [9] | 14+ | TF+Helm | R/M | ○ | × |
| ROBODUCK [5] | 1 | TF+VMs | –/S* | ○ | × |
| FUZZINGBRAIN [44] | 4 | TF+K8s | –/–* | ○ | × |
| ARTIPHISHELL [7] | 53 | TF+Helm | RMQ/PG+N* | △ | × |
| BUGBUSTER [8] | 16 | TF+Helm | RMQ/PG+R* | × | × |
| LACROSSE [28] | 3+ | TF+VMs | RMQ/– | × | × |

TF = Terraform, K8s = Kubernetes, Ka = Kafka, RMQ = RabbitMQ. R = Redis, PG = PostgreSQL, M = MongoDB, S = SQLite, N = Neo4j. * Uses LiteLLM [6] for LLM proxy and routing. ○ Post-competition standalone version available. △ Post-competition local deployment guide available.

### 2.2. From Competition to Deployment

To assess whether AIxCC CRSs can be deployed outside the competition, we analyzed the open-sourced repositories and deployment artifacts of all seven finalists [5, 7–9, 24, 28, 44]. We find three recurring barriers: *infrastructure duplication*, *cloud lock-in*, and *monolithic design*. These findings align with Zhang *et al*. [54], who systematize the techniques of all seven AIxCC finalists and conclude that "the real bottleneck is not technique capability but robust integration into autonomous systems." The next sections show where deployment breaks down in practice.

### 2.3. Barrier 1: Infrastructure Duplication

Table 1 summarizes the deployment characteristics of each team, where *Comp.* denotes the number of independently built container images. The *Infra* and *Middleware* columns show that teams selected different tools yet converged on similar platform roles: container orchestration via Terraform with Kubernetes[26] or Helm[22], and coordination backends such as Kafka[2], RabbitMQ[40], Redis[41], and PostgreSQL[38]. Five of seven teams deployed LiteLLM [6] as their LLM gateway.

The repositories reveal further overlap not visible in the table: each team independently built LLM budget tracking, cost enforcement, and model routing logic on top of its proxy, as well as test environments for applying patches, rebuilding targets, and validating PoVs. As noted in §2.1, LLM budget control is a shared requirement, yet each team built this logic independently. The barrier is not tool diversity itself, but the repeated integration effort: overlapping infrastructure roles reimplemented across all seven teams. This duplication extends beyond initial development: as LLM provider APIs and cloud platforms evolve, each team must independently update its proxy integration, cost-tracking logic, and deployment scripts.

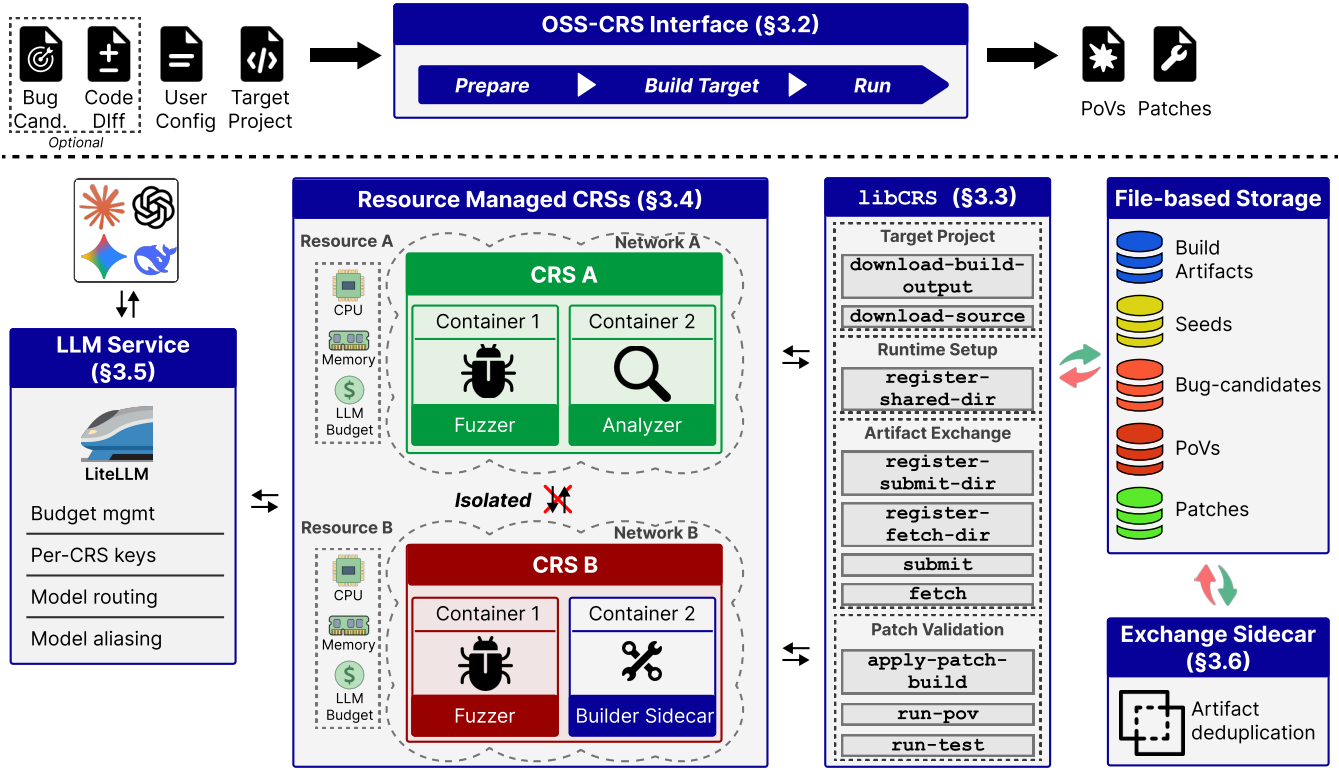

**Figure 1:** Architecture overview of OSS-CRS. Users provide a target project and configuration (optionally with code diffs or bug candidates) and receive PoVs and patches as outputs through the three-phase interface (*prepare*, *build-target*, *run*). CRSs run in resource-managed containers with isolated networks and interact with the platform through `libCRS`. All LLM calls are routed through the LiteLLM proxy, which handles model routing and per-CRS budget enforcement. The exchange sidecar deduplicates artifacts in file-based storage and synchronizes them across all CRSs.

### 2.4. Barrier 2: Cloud Lock-in

The AIxCC competition provisioned each team with dedicated Azure virtual machines orchestrated by Kubernetes. That environment has since been shut down, and the released artifacts still target infrastructure that no longer exists. For example, ATLANTIS, the first-place system [24], requires 20+ Azure VMs, 42 TiB of cloud storage, and runtime Azure SDK calls to scale Kubernetes node pools, despite open-sourcing its full codebase.

As the *Local* column of Table 1 shows, only half the teams added local execution support through standalone releases or local deployment guides [5, 7, 9, 44]. These efforts show that cloud decoupling is possible. However, a local environment must still provide the resource controls and isolation that cloud platforms supply: CPU and memory quotas, network separation between CRSs, and LLM budget limits. Without these, multiple CRSs cannot run on a single machine without contention or cost overruns.

### 2.5. Barrier 3: Monolithic Design

Beyond infrastructure and deployment, a structural obstacle remains: every CRS is a monolith with no modular internal interfaces. The seven finalist teams developed distinctive, often complementary techniques: LLM-based input mutation and hybrid fuzzing [24], LLM-first PoV generation [5], grammar-based fuzzing with LLM-generated grammars [7], expertise-driven multi-agent patching [9], diverse LLM strategy ensembles [44], traditional analysis with LLM-augmented patching [8], and multi-LLM workflow coordination [28]. Comparing and combining these techniques would reveal which approaches outperform others and whether ensembling improves overall results.

Yet as the *Composable* column of Table 1 confirms, no CRS exposes interfaces for component-level extraction. If ATLANTIS has a stronger fuzzer and BUTTERCUP has a stronger patcher, a researcher cannot combine them without reimplementing one inside the other. The AIxCC competition evaluated end-to-end system outputs but provided no way to attribute results to individual techniques; even Zhang *et al.*'s survey [54] could describe each team's methods but not compare them experimentally. As long as CRSs remain monolithic, techniques cannot be isolated, evaluated, or transferred beyond their original systems. A composable framework would let researchers mix the strongest components from different teams, run controlled ablation studies, and build on each other's advances rather than rebuilding entire systems from scratch.

# 3. System Design

To address the barriers identified in §2, we propose OSS-CRS, a locally deployable infrastructure for running and combining existing CRS techniques. OSS-CRS is not itself a CRS; it is the platform on which CRSs are developed, deployed, and composed. It provides three capabilities: 1) a unified three-phase execution model and standard interface (libCRS) that remove per-team infrastructure duplication; 2) local execution with resource controls and isolation across CPU, memory, network, and LLM budgets, removing cloud dependencies; and 3) cross-CRS artifact exchange that enables combining techniques from independently developed CRSs without requiring a single monolithic pipeline. The following subsections detail how OSS-CRS realizes these capabilities.

## 3.1. Architecture Overview

Figure 1 illustrates the architecture of OSS-CRS. The OSS-CRS user interface defines a three-phase lifecycle (*prepare*, *build-target*, *run*) that is driven by a single configuration file that deploys CRSs against target projects. CRSs run in isolated Docker containers with dedicated CPU, memory, and LLM budget allocations; separate networks prevent direct inter-CRS communication. Shared infrastructure provides common services: the *LiteLLM proxy* handles model routing and budget enforcement, *file-based storage* persists artifacts (seeds, PoVs, patches, bug-candidates), and the *exchange sidecar* manages artifact flow between CRSs. CRSs interact with the platform through libCRS, which provides APIs for downloading targets, submitting findings, and validating patches.

## 3.2. OSS-CRS Interface

From the user's perspective, OSS-CRS only requires two inputs: a target project with source code and a single configuration file that specifies which CRSs to deploy along with their resource allocations. Optionally, users can also provide code diffs or bug candidates for targeted bug-finding in environments such as CI/CD pipelines. The user then sequentially runs three commands that set up CRSs, compile the target, and launch the analysis campaign. The main output artifacts are discovered bugs as PoVs and patches that fix them.

**Three-phase lifecycle.** OSS-Fuzz combines building and running into a single workflow: build the fuzz target, then run it. OSS-CRS introduces a three-phase model (*prepare*, *build-target*, *run*) that separates CRS setup from target compilation and running, enabling caching and modularity across diverse CRS architectures. The *prepare* phase builds CRS container images and their dependencies. These images contain the CRS's analysis tools but have no knowledge of the target project. A prepared CRS can analyze any compatible target without rebuilding, amortizing setup cost across multiple analysis campaigns. The *build-target* phase constructs images for target compilation and compiles the target project. Separating this phase from *prepare* allows CRSs to construct target-dependent containers while keeping target-independent components cached. The *run* phase launches all CRS containers and executes the analysis campaign. CRSs operate concurrently, exchanging artifacts through a shared directory and submitting and syncing findings via the libCRS interface. Users can specify a timeout or manually terminate the campaign.

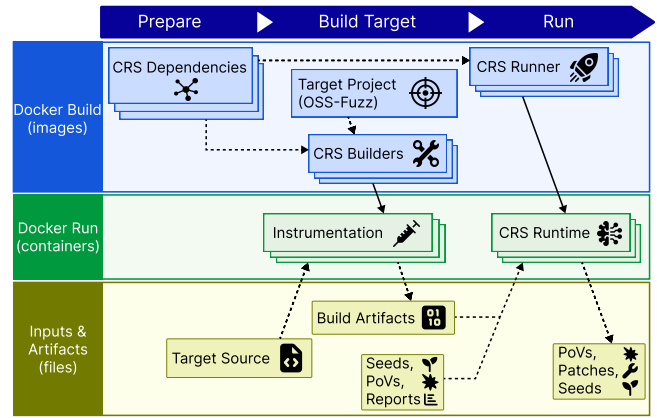

**Figure 2:** Docker workflow across the three operational phases, showing image construction (top), container execution (middle), and file artifact flow (bottom).

Figure 2 summarizes the Docker workflow across three layers. The top layer shows Docker images: *prepare* builds CRS dependency images, *build-target* adds the target project and CRS builder images, and *run* produces the final CRS runner images. The middle layer shows containers: *build-target* runs instrumentation containers that compile the target, and *run* launches CRS runtime containers. The bottom layer tracks file artifacts: target source feeds into build artifacts and seed corpora during *build-target*, and the *run* phase produces PoVs, patches, and new seeds.

**Single configuration.** Users only need to define a single configuration file (crs-compose.yaml; see Listings 4–7) that specifies CRSs, resource allocations, runtime environment, and LLM settings for all three phases. Adding or removing CRSs requires editing this one file, not coordinating multiple configurations. Users can also specify resource constraints without understanding CRS internals. The same file reproduces orchestration and resource settings across different machines.

**Targeted analysis.** By default, CRSs analyze the entire target codebase. For focused analysis (checking whether a recent commit introduces vulnerabilities or fixing a specific reported bug), users can provide targeting metadata that constrains the analysis scope. OSS-CRS accepts targeted inputs through standard channels. A code diff specifies changed code regions, enabling delta analysis between versions; directed fuzzers can focus on changed functions, and patch generators can scope fixes to the modified code. Bug-candidate reports identify specific issues for CRSs to address; OSS-CRS accepts SARIF reports [35] to align with the standard format of existing static analysis tools. Targeted analysis supports CI/CD integration, where CRSs check pull requests rather than the entire codebase.

**TABLE 2:** Representative `libCRS` commands used by CRS developers.

| Category | Command | Purpose |
|---|---|---|
| Build outputs | `submit-build-output` | Publish build artifacts |
| Runtime setup | `register-shared-dir` | Share files within CRS |
| Artifact exchange | `register-submit-dir` | Register for background submission |
| Artifact exchange | `register-fetch-dir` | Register for background fetching |
| Artifact exchange | `submit` | Submit artifact (PoV, seed, patch) |
| Artifact exchange | `fetch` | One-shot fetch of artifacts |
| Patch validation | `apply-patch-build` | Apply patch and rebuild |
| Patch validation | `run-pov` | Run PoV against patched build |
| Patch validation | `run-test` | Run regression tests |

### 3.3. libCRS

`libCRS` is a Python library automatically injected into every CRS container, providing the interface between a CRS's analysis logic and the OSS-CRS infrastructure. A CRS uses the same `libCRS` interface regardless of deployment environment; only the OSS-CRS infrastructure layer changes, not the CRS itself. Table 2 lists the core commands across four categories: *build outputs*; *runtime setup*; *artifact exchange*; and *patch validation*.

**Build outputs.** The *build-target* and *run* phases execute in isolated containers, so build artifacts must be handed off across the phase boundary. Build outputs such as instrumented binaries, source snapshots, and coverage metadata, are published via `submit-build-output` during compilation and retrieved at run time.

**Runtime setup.** A CRS may comprise multiple containers: a fuzzer generating inputs, an analyzer triaging crashes, a patcher synthesizing fixes. The `register-shared-dir` command provides intra-CRS file sharing for corpora, coverage data, and intermediate results, without passing through the inter-CRS exchange.

**Artifact exchange.** CRSs receive external data via `register-fetch-dir` and `fetch`: initial inputs supplied by the operator (seed corpora, reference diffs, bug-candidate reports) and artifacts submitted by other CRSs during ensemble execution. Conversely, CRSs upload their findings (PoVs, seeds, patches) via `register-submit-dir` and `submit`, which flow to other CRSs' fetch directories via the exchange sidecar.

**Patch validation.** Bug-fixing CRSs require a tight edit-compile-test loop to iterate on patch candidates. `libCRS` provides `apply-patch-build`, `run-pov`, and `run-test` for incremental project rebuilding and testing through the builder sidecar.

Figure 3 shows the builder sidecar workflow. During the *build-target* phase, OSS-CRS captures a Docker image snapshot of the fully compiled target along with its build artifacts. In the *run* phase, CRS runtime containers receive the build artifacts and begin analysis. When a CRS generates a candidate patch, it invokes `apply-patch-build`, which sends the diff to the builder sidecar; the sidecar restores the snapshot, applies the patch, and performs an incremental recompilation. The CRS then calls `run-pov` to re-execute the crash-triggering input against the patched binary, and `run-test` to run the project's regression tests. Unlike shared infrastructure services, the builder sidecar runs within the CRS's resource allocation (`cpuset` and `memory_limit`), ensuring that rebuild costs are accounted to the CRS, not the platform.

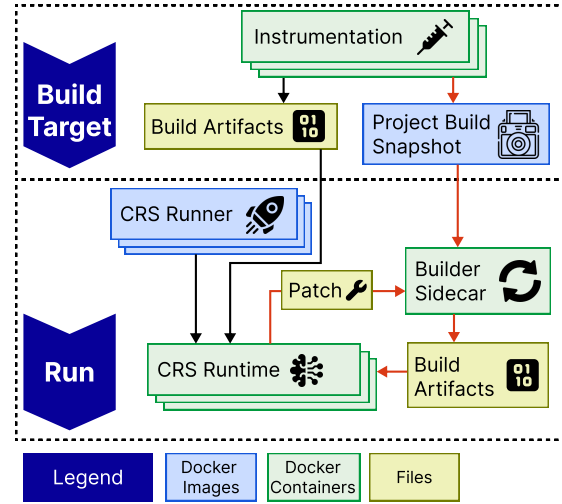

**Figure 3:** Builder sidecar workflow. The sidecar restores a snapshot of the compiled target, applies the patch diff, and performs an incremental rebuild.

### 3.4. Resource Management

Running multiple CRSs simultaneously demands coordinated allocation of heterogeneous resources: CPU cores for fuzzing and analysis, memory for program instrumentation, and LLM API budgets for AI-powered reasoning.

**Compute and memory isolation.** Each CRS declares a `cpuset` string specifying its assigned CPU cores and a `memory` limit enforced via Docker cgroups (`cpuset`, `mem_limit`). Pinning CPUs prevents CRSs from contending for the same cores, while hard memory caps prevent a single CRS from destabilizing others.

**LLM budget as a first-class resource.** Beyond compute resources, OSS-CRS manages LLM API costs as a first-class resource. Each CRS may declare an `llm_budget` in US dollars; the LLM proxy tracks usage against this limit and rejects requests that would exceed it. This prevents cost overruns during long-running campaigns and enables fair comparison between CRSs with different cost profiles: a CRS that achieves the same results within a $50 budget is more efficient than one requiring $500.

**Network isolation.** A shared network introduces potential resource-management issues such as port conflicts between CRS containers and uncontrolled bandwidth consumption that could interfere with per-CRS resource guarantees. OSS-CRS mitigates this by creating separate Docker networks per CRS and restricting all inter-CRS data exchange to the filesystem-based artifact mechanism. This also simplifies access control for shared services like the LLM proxy and ensures fair comparison when benchmarking multiple CRSs side by side.

**Flat Docker architecture.** One approach to porting existing CRSs is to wrap them in Docker-in-Docker (DinD), running the entire system inside a single outer container. However, DinD complicates build performance, resource control, and debugging: Docker layer caches are harder to preserve across builds, host-level cgroup enforcement for individual inner containers is less direct, and debugging nested containers is substantially harder. OSS-CRS instead provides interfaces that let CRS developers build for a flat Docker architecture, where all containers are managed by the host Docker daemon. This enables straightforward per-container resource enforcement, preserves Docker layer caches, and simplifies debugging.

### 3.5. LLM Services

CRSs rely on LLMs for code understanding, patch generation, and vulnerability reasoning. However, different CRSs may prefer different providers and models. OSS-CRS handles these differences with an LLM proxy layer based on LiteLLM [6], an open-source proxy that natively supports multiple provider APIs and model aliasing.

**Unified endpoint.** All CRSs make standard OpenAI-compatible API calls to a single proxy that routes requests to the appropriate backend such as OpenAI, Anthropic, Google, or self-hosted inference servers. A CRS only needs to implement one API client, regardless of which provider the user ultimately configures.

**Model aliasing.** CRSs reference models by logical names (*e.g.*, `claude-sonnet` or `gpt-4o`) rather than provider-specific identifiers. The proxy maps these aliases to concrete provider endpoints, so swapping providers requires only a configuration change, not CRS code modifications. Operators define model mappings in a YAML configuration file (see Listings 8 and 9). Each entry specifies a logical `model_name` (what CRSs request) and provider-specific parameters including the actual model identifier, API credentials, and optional custom endpoints for self-hosted or Azure deployments. At validation time, OSS-CRS checks that each CRS's `required_llms` are available in the operator's configured model list.

**Per-CRS budget enforcement.** OSS-CRS generates a unique API key per CRS at campaign startup, each associated with its budget. The proxy tracks cumulative costs and rejects requests when the budget is exhausted. This per-CRS keying also enables fine-grained usage tracking such as which CRS made each request, what model was used, and how many tokens were consumed. When a CRS exhausts its budget, subsequent requests are rejected. The LiteLLM proxy is deployed as a service in the OSS-CRS infrastructure, and its lifetime is coupled with the run phase.

**Deployment modes.** OSS-CRS supports three deployment modes. In *internal mode*, OSS-CRS manages LiteLLM, and a key-generation sidecar; this is the default for standalone deployments. In *external mode*, operators provide an existing LLM proxy endpoint and key, useful when an organization already runs centralized LLM infrastructure; per-CRS budget enforcement depends on the external proxy's capabilities. In *disabled mode*, OSS-CRS performs no LLM setup, for CRSs that do not use LLMs.

### 3.6. Artifact Exchange

Running multiple CRSs simultaneously enables complementary workflows; a fuzzer can discover crashes while a separate patcher generates fixes. Our evaluation demonstrates that cross-CRS artifact flow works in practice (§4); quantifying performance gains over single-CRS runs remains future work. OSS-CRS enables this *ensemble execution* (see Listing 7) through a filesystem-based exchange mechanism that requires no direct communication between CRSs.

**Exchange model.** All inter-CRS coordination flows through a shared *exchange directory*. Each CRS writes artifacts to its private submit directory, and reads from a shared fetch directory that mirrors the exchange. The exchange sidecar synchronizes per-CRS submit directories into a shared exchange exposed through each CRS's fetch directory. The exchange organizes artifacts by type: `seeds` for fuzzing inputs, `povs` for crash-triggering inputs, `patches` for proposed fixes, and `bug-candidates` for triage reports. CRS-internal data (e.g. coverage maps, model weights, intermediate analysis state) remains private to each CRS and is not shared through the exchange. Artifacts are stored under content-hash filenames, ensuring that duplicate discoveries from multiple fuzzers appear exactly once. This hash-based deduplication is the baseline strategy. Because the exchange sidecar is a shared OSS-CRS service rather than library code launched in a CRS container, we can extend it with more advanced logic without modifying individual CRSs: coverage-based deduplication that keeps only seeds increasing overall coverage, or stack-trace-based deduplication that groups PoVs by crash signature to reduce redundant triage.

**Coordination without direct communication.** CRSs coordinate implicitly through the artifacts they exchange, without direct messaging or task assignment. Each CRS polls for new artifacts at its own pace and decides independently what work to perform. This design provides fault isolation. If one CRS crashes, others continue operating on previously shared artifacts. Resource exhaustion in one CRS does not cascade to others, since each runs in an isolated container with independent resource allocations. This file sharing-based model also simplifies deployment: operators can add or remove CRSs across campaigns without reconfiguring communication channels within CRSs.

TABLE 3: CRSs integrated into OSS-CRS.

| CRS | Type | Languages | LLM |
|---|---|---|---|
| CRS-LIBFUZZER | Bug-finding | C/C++ | None |
| ATLANTIS-C | Bug-finding | C/C++ | ✓ |
| ATLANTIS-JAVA | Bug-finding | Java | ✓ |
| ATLANTIS-MULTILANG | Bug-finding | C/C++/Java | ✓ |
| CLAUDECODE | Bug-fixing | C/C++/Java | ✓ |

TABLE 4: Environment variables injected into CRS containers.

| Variable | Purpose |
|---|---|
| OSS_CRS_TARGET | Target project name |
| OSS_CRS_TARGET_HARNESS | Harness binary name |
| OSS_CRS_NAME | CRS name (for service discovery) |
| OSS_CRS_CPUSET | Allocated CPU cores (*e.g.*, 4-7) |
| OSS_CRS_MEMORY_LIMIT | Memory limit (*e.g.*, 16G) |
| OSS_CRS_LLM_API_URL | LLM proxy endpoint |
| OSS_CRS_LLM_API_KEY | Per-CRS API key for budget enforcement |

## 4. CRS Integration

The goal of OSS-CRS is to provide a modular framework where researchers can study, compare, and compose the distinctive techniques each team developed. As a first phase of validation, we integrated five CRSs that span the spectrum from traditional fuzzing to LLM-powered multi-language analysis (Table 3). The framework is open to the community to port additional systems. Our primary validation focuses on ATLANTIS for two reasons: ❶ it is the best-performing AIxCC finalist, and ❷ it provides no support for local execution outside the competition's cloud infrastructure. The remaining CRSs validate interface usability and ensemble mechanics.

### 4.1. Integration Requirements

CRS developers integrate their system into OSS-CRS through a `crs.yaml` manifest with three phase-aligned sections.

**prepare_phase** builds the CRS's own container images (*e.g.*, fuzzers, analyzers, LLM agents) that are target-independent and reusable across projects.

**target_build_phase** declares build steps that require knowledge of the target: custom instrumentation passes, sanitizer-specific compilations, or snapshot captures for incremental patch validation. In the AIxCC competition, all seven teams independently implemented their own build and patch-compilation pipelines: ATLANTIS built a `cp_manager` with Docker-in-Docker, BUTTERCUP used a dedicated `build-bot` service, and ARTIPHISHELL deployed separate `patchery` and `patch-validation-testing` components [7, 9, 24]. With OSS-CRS, CRS developers extract these custom build passes into `target_build_phase` declarations.

**crs_run_phase** defines the runtime modules as a set of named containers, each with its own Dockerfile. A multi-component CRS (*e.g.*, a fuzzer, an analyzer, and a builder sidecar) declares one entry per component, and OSS-CRS launches them together with shared networking and resource limits.

OSS-CRS also injects environment variables into every CRS container at runtime (Table 4). These expose the target name, resource allocation, and LLM proxy credentials, so CRSs can adapt without hard-coded assumptions. For instance, a fuzzer can spawn one worker per core.

```bash
#!/bin/bash
set -e
compile
libCRS submit-build-output $OUT build
```

**Figure 4:** Build script for CRS-LIBFUZZER.

### 4.2. Baseline: CRS-LIBFUZZER

The simplest integrated CRS is a thin wrapper around libFuzzer [39], the coverage-guided fuzzer used by OSS-Fuzz. It serves as a baseline: it requires no LLM, uses a single container, and exercises the minimal OSS-CRS interface, *i.e.*, prepare, build-target, and run, with nothing beyond standard fuzzing. Its build script (Figure 4) is two lines: `compile` the target, then hand over artifacts to the run phase. This two-line pattern is the minimal build contract; CRSs can extend it freely by adding custom compiler passes, instrumenting with different sanitizers, or running static analyzers, while OSS-CRS handles container orchestration and artifact transfer. The complete `crs.yaml` and `crs-compose.yaml` for CRS-LIBFUZZER appear in Listing 1 and Listing 4.

### 4.3. Porting ATLANTIS

ATLANTIS, the first-place AIxCC system [24], is among the most complex CRSs: it comprises multiple sub-systems (three independent bug-finding systems), uses multiple instrumentation passes (coverage, AddressSanitizer [43], and uninstrumented builds), and combines diverse fuzzing engines with multiple LLM agents.

**Porting process.** Converting ATLANTIS to the OSS-CRS interface required four categories of changes. ❶ *Manifest creation.* We wrote `crs.yaml` manifests for each sub-system, declaring its supported languages, required LLM models, capabilities, and prepare/build/run phase definitions. ❷ *Artifact submission.* Competition-specific API usage (*e.g.*, sending PoVs) was replaced with libCRS commands (`register-submit-dir`, `submit`). ❸ *Project building.* Each ATLANTIS sub-system had its own way of adding CRS tooling to the target project image: ATLANTIS-MULTILANG rebuilt the project image with a custom base image and ATLANTIS-C mounted tooling into the project container and overwrote the compiler environment variable. Under OSS-CRS, all sub-systems follow the same standard: providing a `builder.Dockerfile` that receives the target project base image as a build argument, unifying the build workflow.

❹ *Image caching via prepare phase.* In the competition environment, CRS images are prebuilt into registries. However, for rapid local CRS development, OSS-CRS supports rebuilding CRS images through its infrastructure. As such, each ATLANTIS sub-system needed to optimize its build process by partitioning some images into the prepare phase.

**Three sub-systems ported.** We ported three of ATLANTIS's bug-finding sub-systems as independent CRSs. ATLANTIS-MULTILANG handles C/C++/Java targets using microservices-based fuzzing driven by directed fuzzing and LLM agents. ATLANTIS-JAVA targets Java projects with Jazzer [23] sinkpoint-based fuzzing, ATLANTIS-C targets C/C++ projects with a custom libAFL [17] fuzzer, and both ATLANTIS-JAVA and ATLANTIS-C use a high-throughput agentic seed generator. Each sub-system has its own `crs.yaml` manifest, its own target build process, and runs in its own containers. The configuration files for ATLANTIS-MULTILANG appear in Listing 2 and Listing 5.

**What changed, what was preserved.** The original ATLANTIS requires over 20 Azure VMs across multiple Kubernetes node pools, three separate LiteLLM proxy instances, and dynamic node scaling via Azure SDK calls (§2.4). Porting replaced all Kubernetes orchestration with a flat Docker architecture, collapsed the three LiteLLM proxies into the single OSS-CRS proxy, and substituted the competition scoring API with `libCRS`'s artifact submission. Each sub-system also shed auxiliary components: ATLANTIS-MULTILANG and ATLANTIS-JAVA dropped their concolic execution engines, and ATLANTIS-C dropped its backup fuzzing engines and harness scheduler. All primary bug-finding logic, including fuzzers, LLM agents, and analysis pipelines, was preserved with minimal modification, confirming that the core analysis techniques are independent of the deployment infrastructure.

### 4.4. CLAUDECODE: LLM-Based Patch Generation

To demonstrate bug-fixing integration, we developed CLAUDECODE, a CRS that uses an LLM agent to analyze crash traces and source code for vulnerabilities discovered by other CRSs in order to generate patches.

**Builder sidecar in action.** CLAUDECODE validates each candidate through `libCRS`'s three-step patch validation pipeline: `apply-patch-build` sends the unified diff to a builder sidecar, which applies the patch and performs an incremental rebuild from a pre-captured build snapshot; `run-pov` re-executes the crash-triggering input against the patched binary to confirm the vulnerability is resolved; `run-test` runs the project's regression tests to check for functional regressions. Because the builder sidecar handles all build-system complexity (such as Makefiles), CLAUDECODE's implementation contains no build logic. The CRS focuses entirely on LLM-driven patch synthesis, delegating compilation and testing to the framework.

### 5. Zero-day Results

To evaluate whether OSS-CRS enables practical, large-scale vulnerability discovery and fixing on real-world software, we ran the ported ATLANTIS CRS against open-source projects.

**Target selection.** We selected 8 OSS-Fuzz projects spanning C/C++ and Java, chosen to cover a range of project sizes (3 kLoC to 1,960 kLoC), application domains (databases, parsers, network servers, cryptographic libraries), and existing fuzzing maturity (from newly onboarded to heavily fuzzed for years). We did not cherry-pick projects based on expected vulnerability yield.

**Running setup.** All experiments ran on a single machine with 32 CPU cores and 128 GB RAM, running Ubuntu 22.04 with Docker 27. Each campaign allocated 16 cores and 64 GB RAM to the bug-finding CRS, with a 24-hour timeout per target project. LLM budgets were set to $50 per campaign, using a mix of Claude and GPT-4o through the OSS-CRS LLM proxy.

### 5.1. Zero-day Bugs Found

At the time of writing, three have been fixed by upstream maintainers, one confirmed, and six are pending review. The majority are memory-safety bugs in C, but the set also includes logic bugs and undefined-behavior flaws (CWE-476, CWE-674, CWE-681), demonstrating that the CRS finds null-pointer, schema-validation, and numeric-conversion issues, not only fuzzer-class crashes.

We will release three cases in detail from two projects, following a 30-day window after the original fix was made available.

### 5.2. Patches and Disclosure

**Patch generation and validation.** For each vulnerability, OSS-CRS feeds the crash trace, root-cause analysis, and surrounding source code to ATLANTIS, which uses an LLM to generate a minimal candidate patch as a unified diff. Each candidate is then validated automatically through the `libCRS` infrastructure (§3.3): the patch is applied and the project rebuilt, the original PoV is re-executed to confirm the crash no longer triggers, and the project's test suite is run to check for regressions. We then manually reviewed all patches before submitting them to upstream maintainers.

Three examples will be provided after a 30-day post-fix window to illustrate the kinds of patches OSS-CRS produces.

**Disclosure outcomes.** Of the 10 findings, three have been fixed by upstream maintainers, but have not passed a 30-day period since being fixed. There is one confirmed by maintainers but not yet patched. The remaining six are pending initial response. We reported all vulnerabilities following each project's preferred disclosure process, including GitHub Security Advisories and direct email.

# 6. Discussion

## 6.1. Challenges and Lessons

**Docker-in-Docker vs. flat Docker.** Our initial attempt at local portability wrapped each CRS in a Docker-in-Docker (DinD) container. As discussed in §3.4, DinD complicated resource control, build caching, and debugging, so OSS-CRS adopted a flat Docker architecture instead. The practical lesson is that the flat model requires explicit network policies to prevent cross-CRS communication, a tradeoff we found acceptable given the gains in resource visibility and build performance.

**Build-system diversity.** OSS-Fuzz projects use heterogeneous build systems (Make, CMake, Autoconf, Bazel, Meson, and more). During CRS integration, we found that build assumptions baked into the competition environment (pre-installed tool versions, fixed filesystem layouts) frequently broke on different target projects. OSS-CRS mitigates this by building targets through OSS-Fuzz's official build flows, inheriting the build environment that each project's maintainers already support.

**Porting effort.** Adapting ATLANTIS to the OSS-CRS interface required understanding the system's internal architecture and artifact flow without changing the core logic of the modules we ported. The changes fell into four categories: configuration (writing `crs.yaml` manifests), I/O adaptation (replacing competition-specific API calls with `libCRS` commands), build integration (replacing DinD-based patch compilation with `libCRS`'s builder sidecar, which handles the apply-patch, rebuild, and test cycle), and image optimizations using the prepare phase. For the ATLANTIS-MULTILANG port, this integration work required approximately 3 person-days. For ATLANTIS-C, the port took over 4 person-days, as its tighter coupling to competition-specific build pipelines and container orchestration demanded more extensive I/O and build-integration changes. For ATLANTIS-JAVA and CLAUDECODE, integration required approximately 1 person-day each. Across these ports, we modified configuration, orchestration logic, and artifact I/O boundaries, not the core analysis algorithms. In general, porting effort depends on the gap between a CRS's original design and the four categories above: systems with hard-coded competition APIs, DinD-based build flows, non-standard configuration formats, or tightly coupled image assumptions require proportionally more adaptation work.

## 6.2. Limitations and Threats to Validity

Our results demonstrate feasibility: CRS logic can be decoupled from competition infrastructure and applied to real OSS targets with actionable outputs. We note several limitations of OSS-CRS in its current state.

**Single CRS ported.** We validated OSS-CRS by porting one AIxCC system (ATLANTIS). While ATLANTIS, the first-place system [24], is among the most infrastructure-heavy finalists, porting additional CRSs may reveal interface gaps or design assumptions that a single system does not exercise. We excluded `concolic`, ATLANTIS's hybrid fuzzing component, from the current port because its instrumentation introduced compatibility issues across diverse target projects; in the original competition, this module contributed only 1.7% of ATLANTIS's results [24], so we expect minimal impact on bug-finding capability. We are working on porting other finalist CRSs (§6.3).

**Target format assumptions.** OSS-CRS currently targets projects compatible with OSS-Fuzz, which require a standardized build script, a containerized environment, language restrictions, and at least one fuzz harness. Projects not yet onboarded to OSS-Fuzz require harness development, which is outside the scope of the framework.

**Target selection.** Our 8 target projects were selected from OSS-Fuzz's project corpus based on adoption and security relevance.

**LLM non-determinism.** LLM-based CRS components produce non-deterministic outputs, affecting both vulnerability discovery and patch generation. Our results reflect specific model versions and prompt configurations; different models or API versions may yield different findings.

## 6.3. Community and Future Work

Our roadmap focuses on three directions:
- *Cross-CRS technique analysis.* With ensemble execution and cross-CRS artifact exchange in place, we aim to answer two questions: which individual techniques are most effective, and which combinations yield the best results? Seed deduplication and PoV triage are prerequisites for meaningful cross-CRS comparison.
- *CRS benchmarking.* To enable systematic evaluation, we are developing a benchmark suite that lets security practitioners and researchers evaluate and compare CRSs under controlled conditions, analogous to FuzzBench [34] for fuzzers.
- *Broader target coverage.* We plan to apply integrated CRSs to a wider range of open-source projects to discover vulnerabilities at scale.

# 7. Related Work

**Autonomous cyber reasoning.** The Cyber Grand Challenge (CGC) [13, 48] introduced autonomous systems that find and patch vulnerabilities in custom binaries on the DECREE OS. AIxCC [15] extended this to real-world open-source software, producing seven finalist CRSs whose techniques are analyzed by Zhang *et al.* [54]. OSS-CRS differs from both competition platforms: CGC provided a fixed binary format and a scoring API; AIxCC provided Azure infrastructure with competition-specific endpoints. OSS-CRS provides a reusable framework that persists beyond any single competition, targeting OSS-Fuzz's corpus of open-source projects.

**Fuzzing infrastructure.** OSS-Fuzz [3] provides continuous fuzzing for open-source projects but supports only single-container, single-fuzzer execution without bug-fixing or LLM integration (§2.1). FuzzBench [34] evaluates fuzzer performance on standardized benchmarks but does not support multi-component CRS workloads. ARVO [32] reproduces historical OSS-Fuzz vulnerabilities for research but provides no execution framework. Magma [21] and UniFuzz [30] offer ground-truth benchmarks for evaluating fuzzers but focus on single-fuzzer comparison, not multi-technique composition. OSS-CRS complements these tools: it uses OSS-Fuzz project definitions as targets and could integrate with FuzzBench for CRS-level comparison.

**Ensemble and collaborative fuzzing.** Collaborative fuzzing [12, 19, 20, 56] demonstrated that given fixed resources, distributing effort across multiple fuzzers outperforms focusing on any single fuzzer, motivating ensemble approaches. OSS-CRS generalizes this insight from fuzzer ensembles to CRS ensembles that combine heterogeneous techniques, including fuzzing, static analysis, LLM reasoning, and autonomous patching. It coordinates them via unified resource and LLM budget allocation and a filesystem-based exchange mechanism that requires no modification to individual CRSs.

**LLM-powered fuzzing.** LLMs are increasingly used to augment fuzzing. TitanFuzz [16] and Fuzz4All [50] use LLMs to generate test inputs, ChatAFL [33] applies them to protocol fuzzing, and ELFuzz [11] synthesizes entire fuzzers via LLM-driven evolution. HLPFuzz [51] and $G^2$FUZZ [55] leverage LLMs for constraint solving and input generator synthesis, respectively. In a complementary direction, OSS-Fuzz-Gen [31] and SHERPA [27] use LLMs to synthesize fuzz harnesses, automating a key bottleneck in onboarding new targets. These works improve individual fuzzing components; OSS-CRS provides the orchestration layer to compose such LLM-augmented tools into complete CRS pipelines.

**LLM-based vulnerability repair.** Several benchmarks evaluate LLM agents on security tasks: SEC-bench [29], Cybench [53], AutoPatchBench [10], and CVE-Bench [49]. On the technique side, San2Patch [25] automates vulnerability repair from sanitizer logs via tree-of-thought LLM reasoning, and PatchAgent [52] mimics human debugging expertise for practical program repair. The AIxCC competition demonstrated that integrating LLMs into full CRS pipelines, combining code reasoning with fuzzing and program analysis, yields stronger results than applying them in isolation [54]. OSS-CRS provides the infrastructure to study this integration: its LLM proxy tracks per-CRS costs, enabling comparison of LLM-augmented versus traditional techniques under fixed budgets.

## 8. Conclusion

DARPA's AI Cyber Challenge produced seven autonomous CRSs for finding and fixing vulnerabilities, but left them entangled with competition-specific infrastructure. We analyzed all seven finalist CRSs and identified three categories of barriers (infrastructure duplication, cloud lock-in, and monolithic design) that prevent real-world deployment. OSS-CRS addresses these barriers with a standardized CRS interface, resource isolation with LLM budget management, and cross-CRS artifact exchange for composing complementary techniques. By porting the AIxCC champion system and discovering 10 previously unknown bugs across 8 open-source projects, we provide feasibility evidence that competition CRS logic can be decoupled from its original infrastructure and applied to real-world OSS projects. OSS-CRS is available as open source [37], a first step toward broader cross-CRS accessibility and evaluation. We invite the community, including researchers, CRS developers, and security practitioners, to actively develop, compose, and evaluate CRS techniques and help mature OSS-CRS into shared infrastructure for open-source software security.

## Ethical Considerations

OSS-CRS lowers the barrier to automated vulnerability discovery and patching in open-source software. This automation carries inherent dual-use risks, as the same techniques that help defenders find and fix bugs could also help attackers identify exploitable weaknesses. However, we believe the benefits outweigh these risks: our system produces not only proofs of vulnerability but also validated patches, shifting the balance toward defense.

We ran OSS-CRS entirely in isolated Docker containers on local machines without transmitting any data to external services. Also, because OSS-CRS executes all analysis inside containerized environments and operates on fuzzing harnesses originally written for security testing, it poses no risk to running production systems.

All vulnerabilities discovered in this work were reported to upstream maintainers following each project's preferred disclosure process, including GitHub Security Advisories, SourceForge, and direct email. We withheld public disclosure of vulnerability details until maintainers acknowledged the reports. The proof-of-vulnerability inputs and patches are shared with maintainers to facilitate timely remediation. We are actively communicating with project maintainers and responsible parties to ensure all reported issues are addressed.

## Open Science

OSS-CRS, including the framework, all integrated CRSs, and evaluation artifacts, is publicly available at https://github.com/ossf/oss-crs. The repository includes the CRS interface specification, `libCRS` library, configuration templates, and documentation for porting new CRSs.

## Acknowledgments

We thank Younggi Park for the initial design and implementation of the bug-fixing infrastructure in OSS-CRS. We thank Sin Liang Lee, Isaac Hung, Joshua Wang, and Jiho Kim for their contributions to the bug-finding campaign. We also thank Jeff Diecks and the Open Source Security Foundation (OpenSSF) for supporting OSS-CRS as an OpenSSF sandbox project and promoting community outreach.

# Appendix

We present the complete `crs.yaml` and `crs-compose.yaml` configuration files for the three CRSs integrated in §4: CRS-LIBFUZZER (§4.2), ATLANTIS-MULTILANG (§4.3), and CLAUDECODE (§4.4). The `crs.yaml` manifest declares each CRS's capabilities, supported languages, required LLM models, and build/run phase definitions, while `crs-compose.yaml` specifies resource allocations and runtime settings for a given deployment, including an ensemble example that combines multiple CRSs in a single run. We also include the LiteLLM proxy configurations used for multi-model routing (§3.5): one for ATLANTIS-MULTILANG, which routes directly to upstream providers, and one for CLAUDECODE, which proxies requests to an external LiteLLM instance.

**Listing 1:** `crs.yaml` for CRS-LIBFUZZER

```yaml
name: crs-libfuzzer
type:
  - bug-finding
version: 1.0.0
docker_registry: ghcr.io/oss-crs/crs-libfuzzer

prepare_phase:
  hcl: oss-crs/docker-bake.hcl

target_build_phase:
  - name: build
    dockerfile: oss-crs/dockerfiles/builder.Dockerfile
    outputs:
      - build

crs_run_phase:
  fuzzer:
    dockerfile: oss-crs/dockerfiles/runner.Dockerfile

supported_target:
  mode:
    - full
    - delta
  language:
    - c
  sanitizer:
    - address
  architecture:
    - x86_64
```

```yaml
1  name: atlantis-multilang-wo-concolic
2  type:
3    - bug-finding
4  version: 1.0.0
5  docker_registry: ghcr.io/oss-crs/atlantis-multilang-wo-concolic
6
7  prepare_phase:
8    hcl: oss-crs/docker-bake.hcl
9
10 target_build_phase:
11   - name: uniafl-build
12     dockerfile: oss-crs/dockerfiles/builder.Dockerfile
13     additional_env:
14       BUILD_TYPE: uniafl
15     outputs:
16       - uniafl/build
17       - uniafl/src
18       - uniafl/project
19   - name: coverage-build
20     dockerfile: oss-crs/dockerfiles/builder.Dockerfile
21     additional_env:
22       BUILD_TYPE: coverage
23     outputs:
24       - coverage/build
25   - name: lsp-build
26     dockerfile: oss-crs/dockerfiles/lsp_builder.Dockerfile
27     outputs:
28       - lsp/compile_commands.json
29   - name: create-config
30     dockerfile: oss-crs/dockerfiles/multilang.Dockerfile
31     additional_env:
32       RUN_TYPE: CREATE_CONFIG
33     outputs:
34       - uniafl/config.yaml
35
36 crs_run_phase:
37   multilang:
38     dockerfile: oss-crs/dockerfiles/multilang.Dockerfile
39     additional_env:
40       # Input generators for fuzzing (comma-separated)
41       # Available options:
42       #   - given_fuzzer: Use provided seed corpus (default)
43       #   - testlang_input_gen: Test language-based input generation
44       #   - dict_input_gen: Dictionary-based input generation
45       #   - mlla: Multi-language LLM agent for input generation
46       CRS_INPUT_GENS: given_fuzzer,mlla,testlang_input_gen,dict_input_gen
47   redis:
48     dockerfile: oss-crs/dockerfiles/redis.Dockerfile
49   init_codeindexer:
50     dockerfile: oss-crs/dockerfiles/multilang.Dockerfile
51     additional_env:
52       RUN_TYPE: INIT_CODEINDEXER
53   joern:
54     dockerfile: oss-crs/dockerfiles/joern.Dockerfile
55   lsp:
56     dockerfile: oss-crs/dockerfiles/lsp.Dockerfile
57
58 supported_target:
59   mode:
60     - full
61     - delta
62   language:
63     - c
64   sanitizer:
65     - address
66   architecture:
67     - x86_64
68
69 required_llms:
70   - o4-mini
71   - gpt-4o
72   - gpt-4o-mini
73   - gpt-4.1
74   - gpt-4.1-mini
75   - claude-sonnet-4-20250514
76   - claude-opus-4-20250514
77   - claude-3-7-sonnet-20250219
78   - claude-3-5-haiku-20241022
79   - gemini-2.5-pro
```

Listing 2: `crs.yaml` for ATLANTIS-MULTILANG

```yaml
 1 name: crs-claude-code
 2 type:
 3   - bug-fixing
 4 version: 1.0.0
 5 docker_registry: ghcr.io/oss-crs/crs-claude-code
 6
 7 prepare_phase:
 8   hcl: oss-crs/docker-bake.hcl
 9
10 target_build_phase:
11   - name: default-build
12     dockerfile: oss-crs/builder.Dockerfile
13     outputs:
14       - build
15       - src
16   - name: inc-builder-asan
17     snapshot: true
18     dockerfile: oss-crs-infra:default-builder
19     additional_env:
20       SANITIZER: address
21
22 crs_run_phase:
23   patcher:
24     dockerfile: oss-crs/patcher.Dockerfile
25     additional_env:
26       # Set via compose yaml additional_env.
27       CRS_AGENT: claude_code
28       # ANTHROPIC_MODEL: claude-sonnet-4-5-20250929
29       # CLAUDE_CODE_SUBAGENT_MODEL: claude-sonnet-4-5-20250929
30       # ANTHROPIC_DEFAULT_OPUS_MODEL: claude-sonnet-4-5-20250929
31       # ANTHROPIC_DEFAULT_SONNET_MODEL: claude-sonnet-4-5-20250929
32       # ANTHROPIC_DEFAULT_HAIKU_MODEL: claude-sonnet-4-5-20250929
33       # AGENT_TIMEOUT: "0"  # seconds, 0 = no limit (default)
34   inc-builder-asan:
35     run_snapshot: true
36     dockerfile: oss-crs-infra:default-builder
37
38 supported_target:
39   mode:
40     - full
41     - delta
42   language:  # language-agnostic - agent edits source, builder sidecar handles compilation
43     - c
44     - c++
45     - jvm
46   sanitizer:
47     - address
48     - undefined
49   architecture:
50     - x86_64
51
52 required_llms:
53   - claude-opus-4-6
54   - claude-sonnet-4-6
55   - claude-haiku-4-5-20251001
56 #  - claude-opus-4-5-20251101
57 #  - claude-sonnet-4-5-20250929
58 #  - claude-haiku-4-5-20251001
```

Listing 3: `crs.yaml` for CLAUDECODE

```yaml
 1 # --- General Settings ------------------------------------------------
 2 run_env: local
 3 docker_registry: local
 4
 5 # --- Infrastructure --------------------------------------------------
 6 oss_crs_infra:
 7   cpuset: "0-3"
 8   memory: "16G"
 9
10 # --- CRS (crs-libfuzzer) ---------------------------------------------
11 # Pure fuzzer - no LLM or builder sidecar needed.
12 crs-libfuzzer:
13   cpuset: "4-7"
14   memory: "16G"
```

Listing 4: `crs-compose.yaml` for CRS-LIBFUZZER

```yaml
# --- General Settings -----------------------------------------
run_env: local
docker_registry: local

# --- Infrastructure -------------------------------------------
oss_crs_infra:
  cpuset: "0-3"
  memory: "16G"

# --- CRS (atlantis-multilang-wo-concolic) ---------------------
atlantis-multilang-wo-concolic:
  cpuset: "4-8"
  memory: "16G"
  llm_budget: 100

# --- LLM Configuration ----------------------------------------
llm_config:
  litellm:
    mode: internal
    internal:
      config_path: ./example/atlantis-multilang-wo-concolic/litellm-config.yaml
```

**Listing 5:** `crs-compose.yaml` for ATLANTIS-MULTILANG

```yaml
# --- General Settings -----------------------------------------
run_env: local
docker_registry: local

# --- Infrastructure -------------------------------------------
oss_crs_infra:
  cpuset: "0-1"
  memory: "8G"

# --- CRS (crs-claude-code) ------------------------------------
# Builder sidecars are declared in crs.yaml (snapshot: true / run_snapshot: true)
# and handled automatically by the framework - no separate entry needed.
crs-claude-code:
  cpuset: "2-7"
  memory: "16G"
  llm_budget: 10
  additional_env:
    # Override CRS defaults here. Available models:
    #   claude-opus-4-6, claude-opus-4-5-20251101, claude-opus-4-1-20250805,
    #   claude-sonnet-4-5-20250929, claude-sonnet-4-20250514, claude-haiku-4-5-20251001
    CRS_AGENT: claude_code
    ANTHROPIC_MODEL: claude-opus-4-6
    # CLAUDE_CODE_SUBAGENT_MODEL: claude-sonnet-4-5-20250929
    # ANTHROPIC_DEFAULT_OPUS_MODEL: claude-sonnet-4-5-20250929
    # ANTHROPIC_DEFAULT_SONNET_MODEL: claude-sonnet-4-5-20250929
    # ANTHROPIC_DEFAULT_HAIKU_MODEL: claude-sonnet-4-5-20250929
    # AGENT_TIMEOUT: "3600"                    # Optional: seconds, 0 = no limit (default)

# --- LLM Configuration ----------------------------------------
llm_config:
  litellm:
    mode: internal
    internal:
      config_path: ./example/crs-claude-code/litellm-config.yaml
```

**Listing 6:** `crs-compose.yaml` for CLAUDECODE

```yaml
# --- General Settings ------------------------------------------------
run_env: local
docker_registry: local

# --- Infrastructure --------------------------------------------------
oss_crs_infra:
  cpuset: "0-3"
  memory: "16G"

# --- CRS (crs-libfuzzer) ---------------------------------------------
crs-libfuzzer:
  cpuset: "8-11"
  memory: "16G"

# --- CRS (atlantis-multilang-wo-concolic) ----------------------------
atlantis-multilang-wo-concolic:
  cpuset: "4-7"
  memory: "16G"
  llm_budget: 100

# --- LLM Configuration -----------------------------------------------
llm_config:
  litellm:
    mode: internal
    internal:
      config_path: ./example/atlantis-multilang-wo-concolic/litellm-config.yaml
```

Listing 7: `crs-compose.yaml` for ensemble deployment

```yaml
model_list:
  ######################################
  # OPENAI API
  ######################################
  - model_name: o4-mini
    litellm_params:
      model: openai/o4-mini
      api_key: os.environ/OPENAI_API_KEY

  - model_name: gpt-4o
    litellm_params:
      model: openai/gpt-4o
      api_key: os.environ/OPENAI_API_KEY

  - model_name: gpt-4o-mini
    litellm_params:
      model: openai/gpt-4o-mini
      api_key: os.environ/OPENAI_API_KEY

  - model_name: gpt-4.1
    litellm_params:
      model: openai/gpt-4.1
      api_key: os.environ/OPENAI_API_KEY

  - model_name: gpt-4.1-mini
    litellm_params:
      model: openai/gpt-4.1-mini
      api_key: os.environ/OPENAI_API_KEY

  ######################################
  # ANTHROPIC API
  ######################################
  - model_name: claude-3-7-sonnet-20250219
    litellm_params:
      model: anthropic/claude-3-7-sonnet-20250219
      api_key: os.environ/ANTHROPIC_API_KEY

  - model_name: claude-sonnet-4-20250514
    litellm_params:
      model: anthropic/claude-sonnet-4-20250514
      api_key: os.environ/ANTHROPIC_API_KEY

  - model_name: claude-opus-4-20250514
    litellm_params:
      model: anthropic/claude-opus-4-20250514
      api_key: os.environ/ANTHROPIC_API_KEY

  - model_name: claude-3-5-haiku-20241022
    litellm_params:
      model: anthropic/claude-3-5-haiku-20241022
      api_key: os.environ/ANTHROPIC_API_KEY

  ######################################
  # GEMINI API
  ######################################
  - model_name: gemini-2.5-pro
    litellm_params:
      model: gemini/gemini-2.5-pro
      api_key: os.environ/GEMINI_API_KEY
```

**Listing 8:** LiteLLM proxy configuration for ATLANTIS-MULTILANG (§3.5)

```
 1  model_list:
 2  #####################################
 3  # ANTHROPIC API (via external LiteLLM)
 4  # model_name = what Claude Code sends in API requests (bare names)
 5  # litellm_params.model = how LiteLLM routes to the provider
 6  #####################################
 7  - model_name: claude-opus-4-6
 8    litellm_params:
 9      model: anthropic/claude-opus-4-6
10      api_base: os.environ/EXTERNAL_LITELLM_API_BASE
11      api_key: os.environ/EXTERNAL_LITELLM_API_KEY
12
13  - model_name: claude-sonnet-4-6
14    litellm_params:
15      model: anthropic/claude-sonnet-4-6
16      api_base: os.environ/EXTERNAL_LITELLM_API_BASE
17      api_key: os.environ/EXTERNAL_LITELLM_API_KEY
18
19  - model_name: claude-haiku-4-5-20251001
20    litellm_params:
21      model: anthropic/claude-haiku-4-5-20251001
22      api_base: os.environ/EXTERNAL_LITELLM_API_BASE
23      api_key: os.environ/EXTERNAL_LITELLM_API_KEY
24
25  - model_name: claude-opus-4-5-20251101
26    litellm_params:
27      model: anthropic/claude-opus-4-5-20251101
28      api_base: os.environ/EXTERNAL_LITELLM_API_BASE
29      api_key: os.environ/EXTERNAL_LITELLM_API_KEY
30
31  - model_name: claude-opus-4-1-20250805
32    litellm_params:
33      model: anthropic/claude-opus-4-1-20250805
34      api_base: os.environ/EXTERNAL_LITELLM_API_BASE
35      api_key: os.environ/EXTERNAL_LITELLM_API_KEY
36
37  - model_name: claude-sonnet-4-5-20250929
38    litellm_params:
39      model: anthropic/claude-sonnet-4-5-20250929
40      api_base: os.environ/EXTERNAL_LITELLM_API_BASE
41      api_key: os.environ/EXTERNAL_LITELLM_API_KEY
42
43  - model_name: claude-sonnet-4-20250514
44    litellm_params:
45      model: anthropic/claude-sonnet-4-20250514
46      api_base: os.environ/EXTERNAL_LITELLM_API_BASE
47      api_key: os.environ/EXTERNAL_LITELLM_API_KEY
```

**Listing 9:** LiteLLM proxy configuration for CLAUDECODE, forwarding to an external LiteLLM instance (§3.5)